# Extended-SWIR Photodetection in All-Group IV Core/Shell Nanowires

Lu Luo,[†] Simone Assali,[†] Mahmoud R. M. Atalla,[†] Sebastian Koelling,[†] Anis Attiaoui,[†]

Gérard Daligou,[†] Sara Martí,[‡] J. Arbiol,[‡,§] and Oussama Moutanabbir *,[†]

[†] Department of Engineering Physics, École Polytechnique de Montréal, C. P. 6079, Succ. Centre-Ville, Montréal, Québec H3C 3A7, Canada

[‡] Catalan Institute of Nanoscience and Nanotechnology (ICN2), CSIC and BIST, Campus UAB, Bellaterra, 08193 Barcelona, Catalonia, Spain

[§] ICREA, Pg. Lluís Companys 23, 08010 Barcelona, Catalonia, Spain

**Abstract**

Group IV $Ge_{1-x}Sn_x$ semiconductors hold the premise of enabling broadband silicon-integrated infrared optoelectronics due to their tunable bandgap energy and directness. Herein, we exploit these attributes along with the enhanced lattice strain relaxation in $Ge/Ge_{0.92}Sn_{0.08}$ core-shell nanowire heterostructures to implement highly responsive, room-temperature short-wave infrared nanoscale photodetectors. Atomic-level studies confirm the uniform shell composition and its higher crystallinity with respect to thin films counterparts. The demonstrated $Ge/Ge_{0.92}Sn_{0.08}$ p-type field-effect nanowire transistors exhibit superior optoelectronic properties achieving simultaneously a relatively high mobility, a high ON/OFF ratio, and a high responsivity, in addition to a broadband absorption in the short-wave infrared range. Indeed, the reduced bandgap of the $Ge_{0.92}Sn_{0.08}$ shell yields an extended cutoff wavelength of 2.1 μm, with a room-temperature responsivity reaching 2.7 A/W at 1550 nm. These results highlight the potential of $Ge/Ge_{1-x}Sn_x$ core/shell nanowires as silicon-compatible building blocks for nanoscale integrated infrared photonics.

Group IV semiconductor $Ge_{1-x}Sn_x$ alloys have been at the core of the relentless course toward monolithically integrated infrared optoelectronics and photonics on silicon platforms.[1-3] By controlling the Sn content and lattice strain, the bandgap of $Ge_{1-x}Sn_x$ can be tuned in the 1.4-8 µm range,[4-8] thus enabling key building blocks for potential short-wave infrared (SWIR) and mid-wave infrared (MWIR) applications, including data communication, night vision, thermal imaging, and chemical and biological sensing. Moreover, the hole mobility is predicted to be higher in $Ge_{1-x}Sn_x$ than Ge, thus motivating the use of GeSn-based materials as p-channel metal-oxide-semiconductor field effect transistors (MOSFETs).[9] In addition, $Ge_{1-x}Sn_x$ alloys are compatible with the mainstream silicon-based complementary metal oxide semiconductor (CMOS) processing, which can be a distinct advantage over mainstream compound semiconductors. However, due to the large difference in lattice parameter (~14.7%) between Ge and Sn, epitaxially-grown $Ge_{1-x}Sn_x$ thin films on Ge-buffered silicon substrates are highly compressively strained, which increases the Sn content needed to reach the direct bandgap.[8, 10] Moreover, the strain relaxation during growth results in an undesired formation of defects and dislocations, which degrades device performance.

A promising paradigm to minimize strain in $Ge_{1-x}Sn_x$ is to grow the material as nanowires (NWs) or NW heterostructures,[11-15] where the elastic strain can be relaxed at the free sidewall facets,[16] reducing structural defects and consequently improving the device performance. Despite these advantages, the literature on $Ge_{1-x}Sn_x$-based NWs remains rather scarce,[17-20] with only few attempts so far to employ them in nanoscale devices.[21, 22] As a matter of fact, the lack of detailed knowledge on the interplay between their structural and optoelectrical properties has been limiting the optimization of their performance and consequently their use in devices. With this perspective, this work exploits the versatility of core-shell architectures to fabricate metal-semiconductor-metal (M-S-M) back-gated FET Ge/Ge$_{0.92}$Sn$_{0.08}$ core/shell NW photodetector devices and investigate their optical and electrical properties. The room-temperature operation for these nanoscale photodetectors yields a cutoff at an extended wavelength of ~2.1 µm compared to 1.6 µm in Ge. This is related to the reduced bandgap energy due to alloying with Sn along with the enhanced lattice relaxation. The measured responsivity is 2.7±0.3 A/W under 1550 nm illumination (at 1 V) with selectivity to light polarization. Moreover, p-type conductivity with an average hole mobility of 100 cm$^2$V$^{-1}$s$^{-1}$ was measured in these NWs, which is significantly larger than the values obtained for Ge NW reference devices (up to 40 cm$^2$V$^{-1}$s$^{-1}$). Therefore, Ge/Ge$_{0.92}$Sn$_{0.08}$ core/shell NWs enable an all-group IV platform to cover a range of the electromagnetic spectrum that has been heretofore mainly served by III-V NWs.[23-25]

Ge/Ge$_{0.92}$Sn$_{0.08}$ core/shell NWs were grown on Si(111) wafers, in a chemical vapor deposition (CVD) reactor using germane (GeH$_4$) and tin-tetrachloride (SnCl$_4$) as precursor gases. Prior to the growth, a 2 nm-thick Au layer was deposited on the Si substrate, followed by an HF cleaning step before loading the sample into the CVD reactor. The vapor-liquid-solid (VLS) growth of Ge NWs with diameters in the 30-60 nm range was performed at 330 °C. The temperature was then reduced to 310 °C (Ge/Sn ratio in gas phase of 120) for the growth of the Ge$_{0.92}$Sn$_{0.08}$ shell at thicknesses in the 60-100 nm range. The scanning electron micrograph (SEM) in Figure 1a displays Ge/Ge$_{0.92}$Sn$_{0.08}$ core/shell NWs with lengths of 2-3 µm and diameters of 150-260 nm. The tapering in some of the NWs is a result of an enhanced precursor decomposition in the proximity of the Au droplet.[17] We highlight that only NWs with a small tapering (diameter variation less than 25 nm) were selected to fabricate single NW devices to minimize any effect of tapering on the NW device performance. The NW structural properties were investigated using transmission electron microscopy (TEM) and atom probe tomography (APT). The sunburst-like morphology of the $Ge_{0.92}Sn_{0.08}$ shell, with Sn-rich facets and Ge-rich corner facets[16-18, 26] is visible in the cross-sectional high angle annular dark field (HAADF) scanning-TEM (STEM) image (Fig. 1b) and electron energy loss spectroscopy (EELS) related composition maps (Fig. 1d-f). The individual EELS maps for Ge and Sn in the NW cross section are shown in Figure 1d and e, respectively. The resulting compositional map with the two elements is also displayed in Figure 1f. Note that the yellow arrow in Figure 1e indicates the signal from O atoms in the protective oxide layer. The NW Sn composition radial profile is extracted after excluding the protective oxide shell, as indicated by the white dotted arrow in Figure 1f. Moreover, the high-resolution STEM (HRSTEM) and the related fast-Fourier transform (FFT) image in Figure 1c demonstrate the high crystalline quality of as-grown NWs, without the presence of extended defects in the shell (see supplementary information S1). A precise estimation of the composition of the shell is obtained from APT measurements, as exemplified in Figure 1g. The Sn composition profile along the <112> radial direction of the shell was measured by EELS and APT, as shown in Figure 1h. These two measurements show good agreement in the Sn composition of the shell, where a steep transition (~10 nm-wide) to ~5 at.% Sn, followed by a slight graded profile up to ~8 at.% was observed. A rather constant composition of ~2.5 at.% is visible along the <110> direction of the shell. Note that only ~60 % of the cross-section of the NWs is visible in APT due to the limited field-of-view of the instrument[27], however the comparison with the EELS data shows a constant composition of ~8 at.% in the outer region of the shell.

Back-gated single Ge/Ge$_{0.92}$Sn$_{0.08}$ core/shell NWs devices were fabricated after transfer on a SiO$_2$/Si substrate by electron beam lithography (EBL), and Cr and Au were evaporated to form metal contacts. For comparison, single Ge NW devices were also fabricated following the same protocol. A representative SEM image of a Ge/Ge$_{0.92}$Sn$_{0.08}$ NW device is shown in Figure 2a, while the current-voltage source-drain (I$_{ds}$-V$_{ds}$) characteristics acquired on four devices with different dimensions are shown in Figure 2b. By varying the V$_{ds}$ at a floated-gate bias V$_{gs}$, near ohmic behavior is observed in the -0.5–0.5 V range. The difference in the dark current of four different NW devices in Figure 2b could be related to variations in the contact resistance linked to possible residuals of GeSn native oxide on the NW sidewall. The transfer I$_{ds}$-V$_{gs}$ characteristic in Figure 2c for the Ge/Ge$_{0.92}$Sn$_{0.08}$ core/shell NW FET shows p-type conductivity, where the excess hole free carriers could result from vacancy complexes that might be unintentionally formed during the low temperature shell growth, which is similar to what is commonly observed in Ge NW FET.[28, 29] The ON/OFF ratio of ~100 was estimated in Ge/Ge$_{0.92}$Sn$_{0.08}$ core/shell NW FET (Fig. 2c). This value is lower than what reported earlier for VLS-grown Ge$_{0.91}$Sn$_{0.09}$ NWs (~780).[21] This difference may be attributed to the smaller NW diameter (120 nm) used in the latter, which allows a better gate control over the channel, in turn suppressing the off-state current leakage. The output I$_{ds}$-V$_{ds}$ characteristic in Figure 2d exhibits a linear scaling at small V$_{ds}$ (Fig. 2d), which is an indication of the Ohmic behavior at the contact. The field-effect hole mobility ($\mu_h$) was estimated from the transfer curve using the following equation:

$$\mu_h = g_m \cdot \frac{L^2}{C_{ox}} \cdot \frac{1}{V_{ds}}, \quad \text{(Eq. 1)}$$

where $g_m$ is the transconductance, defined as $\frac{\partial I_{ds}}{\partial V_{gs}}$. Gate capacitance $C_{ox}$ is estimated from the finite element method (FEM) (see Supplementary Information S2 for details), and L is the gate channel length. The estimated field-effect hole mobility of single Ge/Ge$_{0.92}$Sn$_{0.08}$ core/shell NW modulated by gate voltage at V$_{ds}$ = -0.1 V is displayed in Figure 2e. Peak mobility as large as 110 cm$^2$V$^{-1}$s$^{-1}$ was recorded for these NWs, with mean mobility averaged over 11 devices of 100±40 cm$^2$V$^{-1}$s$^{-1}$ (Figure 2f). These values are significantly larger than those of VLS-grown Ge$_{0.91}$Sn$_{0.09}$ NWs with diameters in the 100-150 nm range (15 cm$^2$V$^{-1}$s$^{-1}$),[21] and also larger than those of Ge NW FET reference devices with diameters of 40-60 nm (23 cm$^2$V$^{-1}$s$^{-1}$). The VLS-grown Ge$_{0.91}$Sn$_{0.09}$ NWs in Ref. 21 were fabricated at 440 °C using a liquid injection chemical vapor deposition (LICVD) reactor, therefore a direct comparison on materials properties may be irrelevant due to the different growth parameters and structure compared to Ge/Ge$_{0.92}$Sn$_{0.08}$ core/shell NWs. The mobility variation of Ge/Ge$_{0.92}$Sn$_{0.08}$ core/shell NWs is due to the difference in transconductance and gate capacitance of each NW, which are affected by contact resistance and the fluctuation in the geometrical parameters, respectively.[30] The high mobility of the Ge/Ge$_{0.92}$Sn$_{0.08}$ core/shell NWs most likely results from the use of a larger NW diameter and possibly from the reduced effective mass that is predicted in Ge$_{0.92}$Sn$_{0.08}$.[31] The contribution of surface and defects scattering processes to the measured mobility should also be considered,[32] however at this stage it is extremely challenging to decouple these different effects.

To elucidate the room-temperature optical response of single NW devices, the output curves with a floating gate acquired in dark and under a 1550 nm laser illumination were measured, as shown in Figure 3a. The photocurrent is 40 times larger than the dark current, which indicates the high device sensitivity due to the enhanced absorption in the NW geometry.[33, 34] From the photocurrent measurements, the detector responsivity (R) was estimated as:

$$R = \frac{I_{ph}}{P*A} \text{ , (Eq. 2)}$$

where $P$ is the power density and $A$ is the geometrical cross-section area of the NW device. In NWs, the absorption cross-section is larger than their geometrical one,[35] therefore, the $R$ value provided here should be considered as an upper limit. The inset in Figure 3a displays the evolution of the responsivity as a function of drain-source bias. The responsivity shows a significant increase above 0.1 V. This increase slows down at high drain-source voltage, eventually reaching saturation when virtually all the photogenerated carriers are collected by the electrodes.[36, 37] The power-dependent photocurrent measurements displayed in Figure 3b indicates a linear increase with the excitation power, with the exponent $\theta=1.0$, estimated by fitting the data with a power law $I_{ph} \propto P^{\theta}$. The parameter $\theta$ can be correlated to the complex process of electron-hole generation, trapping, and recombination within the semiconductor,[38] and a linear growth of photocurrent will be observed with a moderate light intensity.[39, 40] The responsivity of Ge/Ge$_{0.92}$Sn$_{0.08}$ core/shell NW photodetector is independent of the excitation power density, with a value of 2.7±0.3 A/W at $V_{ds}$ =1 V. For comparison, the Ge NW photodetectors show a responsivity of only ~30 mA/W at $V_{ds}$ =1 V under a power intensity of 0.27 W/mm$^2$ (Supplementary Information S4). It is worth mentioning that the NW dark current and the photoconductivity can be largely affected by the diameters since the depleted space charge layer caused by Fermi-level pinning can extend about 50-100 nm into the bulk material.[41] Therefore, a small diameter NW can be completely depleted, which may explain the low dark and photo currents of Ge NWs compared to thicker Ge/Ge$_{0.92}$Sn$_{0.08}$ core/shell NWs (we noticed a change from a few pA to a few nA). In addition, coupling incident light to leaky-mode resonance (LMR) supported by NWs can enhance the light-matter interaction, as is discussed later in the text. Time-dependent photocurrent measurements (1s pulses) of Ge/Ge$_{0.92}$Sn$_{0.08}$ NWs with V$_{ds}$ = 0.1 V and V$_{gs}$ = 0 V are shown in Figure 3c. The NW photodetector shows rapid switching between the ON and OFF states, with photocurrent and dark current values that remain stable over the measured 10 periods, thus indicating that parasitic capacitance does not significantly affect device operation.

Polarization-dependent photocurrent measurements under 1550 nm illumination performed on single Ge/Ge$_{0.92}$Sn$_{0.08}$ and Ge NW photodetectors (Figure 3d) show enhanced light absorption along the (111) NW growth axis, which is an inherent property of the one-dimensional NW geometry.[42] When the NW diameter is sufficiently smaller than the excitation wavelength,[43, 44] the polarization ratio $\rho$ becomes highly dependent on the dielectric constants of the NW ($\epsilon$) and the surrounding medium ($\epsilon_0$), according to the equation:[45]

$$\rho = \frac{I_{//}-I_{\perp}}{I_{//}+I_{\perp}} = \frac{(\epsilon+\epsilon_0)^2-4\epsilon_0^2}{(\epsilon+\epsilon_0)^2+8\epsilon_0^2} \text{ , (Eq. 3)}$$

where $I_{//}$ and $I_{\perp}$ are the photocurrent generated by light polarized parallel and perpendicular to the NW axis, respectively. Indeed, the experimental polarization ratio of Ge NW at a diameter of 60 nm is 0.9, which agrees with the predicted polarization ratio of 0.96 ($\epsilon_{Ge}$ = 16).

In a Ge/Ge$_{0.92}$Sn$_{0.08}$ NW at a diameter of ~240 nm, a polarization ratio $\rho = 0.3$ is obtained, which is ~3 times lower than the value measured in the thinner Ge NW. In principle, both Ge$_{1-x}$Sn$_x$ shell thickness and Ge core diameter can be precisely controlled to reduce the overall NW diameter and, if needed, further engineer $\rho$, which will enable device optimization for polarization-sensitive infrared light detection using Ge/ Ge$_{1-x}$Sn$_x$ core/shell NWs.

Spectrally resolved photocurrent measurements performed on a Ge/Ge$_{0.92}$Sn$_{0.08}$ core/shell NW (260 nm diameter) and a Ge NW (100 nm diameter) are shown in Figure 4a. A monotonous decrease in responsivity is visible in the Ge NW until reaching the 1.6 μm cutoff wavelength. A strong peak is visible at 1.42 μm in the Ge/Ge$_{0.92}$Sn$_{0.08}$ core/shell NW, followed by a progressive reduction in intensity at longer wavelengths extending beyond the Ge cutoff wavelength. The observed photocurrent peak is associated with the presence of LMR[34, 42] in the Ge/Ge$_{0.92}$Sn$_{0.08}$ NW, as demonstrated by the finite-difference time-domain (FDTD) simulations displayed in Figure 4a (solid lines). Good agreement between the simulated absorption and the measured responsivity is observed for both Ge and Ge/Ge$_{0.92}$Sn$_{0.08}$ NW devices. Due to the smaller diameter of 100 nm in the Ge NW, no LMR peaks are observed above 1.3 μm, whereas in the thicker 260 nm-diameter Ge NW a resonance peak centered at 1.35 μm is predicted (black dashed curve in Fig. 4a). This simulated peak is located at a shorter wavelength compared to the 1.42 μm peak in a Ge/Ge$_{0.92}$Sn$_{0.08}$ NW at the same diameter due to the difference in the refractive index between Ge and Ge$_{0.92}$Sn$_{0.08}$. Note that unpolarized source was set in the FDTD simulation to match the experiments. (Supplementary Information S5) The enhanced responsivity (Fig. 3) of the Ge/Ge$_{0.92}$Sn$_{0.08}$ NW core/shell heterostructure in the 1.55 μm band is a significant advantage over conventional Ge devices, with the position of the resonance peak that can be in principle controlled across the telecommunication range by tailoring the diameter and composition of the $Ge_{1-x}Sn_x$ NWs. The reduced signal-to-noise ratio at wavelengths longer than 2.0 μm makes it challenging to precisely estimate the cutoff wavelength of a single $Ge/Ge_{0.92}Sn_{0.08}$ NW device. To overcome this, a photodetector made of three $Ge/Ge_{0.92}Sn_{0.08}$ core/shell NWs was fabricated, and the room-temperature spectral responsivity ($V_{ds}$=0.5 V) is displayed in Figure 4b. The absorption peak at 1.42 μm is not resolved in the measurements possibly due to the fluctuation in diameter between different NWs. A monotonic decrease in responsivity is obtained in $Ge/Ge_{0.92}Sn_{0.08}$ three-NWs device, and the well-resolved cutoff at 2.10 μm is observed. We note that by considering the variation in compressive strain (<0.2 %) and composition in the shell (<0.5 at.%) originating from the spread in the Ge core diameter, the calculated corresponding shift in the cutoff wavelength is limited to less than ±0.08 μm, which can hardly be detected in Fig. 4. The cutoff wavelength measured from responsivity spectrum is close to the direct bandgap value of 565 meV (2.2 μm) obtained from *k.p* calculations at 300K for an almost fully relaxed $Ge_{0.92}Sn_{0.08}$ material.[46] For a fully strained (112)-oriented $Ge_{0.92}Sn_{0.08}$ shell, the calculated bandgap energy is 712.7 meV (1.76 μm), which is significantly far from the experimental observation thus highlighting the enhanced relaxation in NWs. Also, the measured cutoff wavelength is consistent with the measurements performed on $Ge_{1-x}Sn_x$ thin films[47-49] and NWs[17] at similar compositions. It is worth noticing that the $Ge_{0.92}Sn_{0.08}$ shell will induce tensile strain in the Ge core up to ~1 %,[16] which will shift the absorption edge of Ge up to ~1.7 μm.[50] However, from the experimental analysis, it is not possible to decouple the Ge core and $Ge_{0.92}Sn_{0.08}$ shell absorption edges, as the absorbing volume of the Ge core is significantly lower than that of the $Ge_{0.92}Sn_{0.08}$ shell.

Table 1 summarizes the electrical and optical parameters of the demonstrated NW photodetectors compared to the few available studies on $Ge_{1-x}Sn_x$ -based nanoscale and thin film devices reported to date. The differences in device structures and compositions are also indicated. A distinctive feature of $Ge/Ge_{0.92}Sn_{0.08}$ NWs devices is the ability to simultaneously achieve relatively higher performance in terms of mobility, ON/OFF ratio, responsivity, and broadband absorption at short wave infrared wavelengths. Remarkably, these devices can operate at room temperature at a voltage reaching 1 V with a responsivity of 2.7 A/W at 1550 nm, which is well above what was recently obtained in ferroelectric enhanced dual-NWs ($1.2\times10^{-3}$ A/W at 1 mV).[51] Moreover, the average hole mobility in $Ge/Ge_{0.92}Sn_{0.08}$ NWs is three times higher than the reported one in polycrystalline $Ge_{0.97}Sn_{0.03}$[52] and almost seven times higher than that obtained in VLS-grown $Ge_{0.91}Sn_{0.09}$ NWs.[21]

The performance of these nanoscale devices depends on the structural and optoelectronic properties of the material, which are affected by the growth protocols employed. On the other hand, the core/shell NW geometry is a promising approach to form nanoscale semiconductors with high material quality resulting from the relaxed strain along the radial direction, which can help form heterostructures with limited defects even on lattice-mismatched materials.[17, 18, 53] Therefore, the high optical and electrical performance of the $Ge/Ge_{0.92}Sn_{0.08}$ core/shell NW is attributed to the high crystalline quality and uniform shell composition. Moreover, these NW devices show higher responsivity compared to thin-film photodetectors at a similar composition and a >300 nm-thick active layer ($4\times10^{-3}$ A/W with an electric field of 0.16 kV/cm).[49] While this difference most likely arises from the inherently reduced light absorption in the planar geometry compared to NWs, the higher structural quality of the latter could also play a role in enhancing the device efficiency. Indeed, all $Ge_{1-x}Sn_x$-based nanoscale devices share a relatively low dark current (hundreds of nA range) compared to as-grown $Ge_{1-x}Sn_x$ thin film devices with a much larger device volume, where values up to several mA are commonly measured.[3, 49] It is documented that the presence of defects in $Ge_{1-x}Sn_x$ thin film can be a source of leakage current in planar optoelectronic devices,[1] while the NW quasi-one-dimensional nature strongly helps reduce the amount of defects (Fig. 1c), resulting in a lower dark current.

In summary, we demonstrated room-temperature $Ge/Ge_{0.92}Sn_{0.08}$ core/shell NW photodetectors with an extended cutoff wavelength of 2.1 μm. These NWs allow an enhanced relaxation of the compressive strain and a reduction of growth defects that are inherent to thin films and typically associated with a limited performance. Consequently, the grown $Ge/Ge_{0.92}Sn_{0.08}$ core/shell NWs exhibit a high crystallinity and a uniform Sn content, and single NW p-type FETs fabricated using these core/shell NWs provide a superior optoelectronic performance. A relatively high mobility (100 $cm^2V^{-1}s^{-1}$ in average), a high ON/OFF ratio (~$10^2$), and a high responsivity (2.7 A/W at 1550 nm) are simultaneously achieved, along with a broadband absorption in the SWIR range. Moreover, the $Ge/Ge_{0.92}Sn_{0.08}$ NW devices were also found to outperform Ge NW devices in both carrier mobility and optical response in the 1.55 μm wavelength telecommunication band, where resonance peaks can be controlled by tuning the NW diameter. These results highlight the potential of $Ge/Ge_{1-x}Sn_x$ core/shell NWs for Si-compatible integrated optoelectronics in the extended SWIR wavelength range, which is relevant to a variety of nanoscale sensing and imaging applications.

**METHODS**

**Epitaxial growth.** The VLS-growth of the Ge/Ge$_{0.92}$Sn$_{0.08}$ core/shell NWs was performed in a low-pressure CVD reactor using monogermane (GeH$_4$) and tin tetrachloride (SnCl$_4$) precursors and ultrapure H$_2$ as carrier gas. First, a 2 nm-thick Au layer was deposited on a Si (111) wafer substrate, followed by HF-last cleaning process prior to loading in the CVD reactor. The Ge NWs were then grown using a two-step process, with a nucleation step at 400 °C under GeH$_4$ supply and Ge NW growth at 330°C. Next, the GeH$_4$ supply was removed, and the sample cooled down to 310 °C where the Ge$_{0.92}$Sn$_{0.08}$ shell was grown for 20 minutes using Ge/Sn ratio in the gas phase of 120.

**Characterization.** Aberration corrected (AC) HAADF-STEM was carried out using a FEI Titan probe corrected transmission electron microscope. Electron Energy Loss Spectroscopy in STEM mode (EELS-STEM) compositional maps have been obtained in a Tecnai F20 microscope by using a GATAN QUANTUM filter.

**Device Fabrication and Optoelectronic Measurements.** The Ge/Ge$_{0.92}$Sn$_{0.08}$ NWs were drop-casted onto a highly doped p-type Si substrate with a 100-nm-thick SiO$_2$ layer. Then spin-coat methyl methacrylate (MMA) and poly (methyl methacrylate) (PMMA) on the SiO$_2$/Si substrate contains NWs. EBL (Raith 150) was used to define the drain and source contact. And the E-beam evaporator was used to deposit Cr/Au 20/280 nm. The room temperature optical and electrical measurements of Ge/Ge$_{0.92}$Sn$_{0.08}$ core/shell NW and Ge NW devices were conducted on a probe station connected with a Keithley 4200 A semiconductor analyzer. Light from 1550 nm laser diode was focused by a 10× objective lens before hit the NW devices. Using a knife edge, the diameter of the Gaussain beam spot size is measured as about 53 μm. The spectral photoresponsivity was measured with Bruker Vertex 80 Fourier-transform infrared spectroscopy (FTIR) spectrometer connected with a Zurich Instruments lock-in amplifier which is triggered by an optical chopper. The electrical signal from the lock-in amplifier is fed into the FTIR to get the photocurrent as a function of wavelength, and the power of the NIR light source is calibrated by a reference InGaAs photodiode.

## ASSOCIATED CONTENT

**Supporting Information.**

The Supporting Information is available free of charge at ACS website.

The Supporting Information contains details of the transmission electron microscopy measurements, electrical and optical measurements of Ge NW devices, and additional data that supports the findings reported in the main manuscript.

**Corresponding Authors**

*E-mail: oussama.moutanabbir@polymtl.ca;

**Notes**

The authors declare no competing financial interest.

**Acknowledgements**

The authors thank J. Bouchard for the technical support with the CVD system and B. Baloukas for support with the SE measurement. O.M. acknowledges support from NSERC Canada (Discovery, SPG, and CRD Grants), Canada Research Chairs, Canada Foundation for Innovation, Mitacs, PRIMA Québec, and Defence Canada (Innovation for Defence Excellence and Security, IDEaS). L.L acknowledges support from China Scholarship Council (CSC). S.A. acknowledges support from Fonds de recherche du Québec-Nature et technologies (FRQNT, PBEEE scholarship). ICN2 acknowledges funding from Generalitat de Catalunya 2017 SGR 327. ICN2 is supported by the Severo Ochoa program from Spanish MINECO (Grant No. SEV-2017-0706) and is funded by the CERCA Programme / Generalitat de Catalunya. S.M. and J.A. acknowledge the use of instrumentation as well as the technical advice provided by the National Facility ELECMI ICTS, node "Laboratorio de Microscopías Avanzadas" at University of Zaragoza. S.M. and J.A. also acknowledge support from CSIC Research Platform on Quantum Technologies PTI-001.

**Supporting Information**

Additional information on the structural characterization of the NWs, optoelectronic measurements, and FDTD absorption simulation.
This material is available free of charge at http://pubs.acs.org.

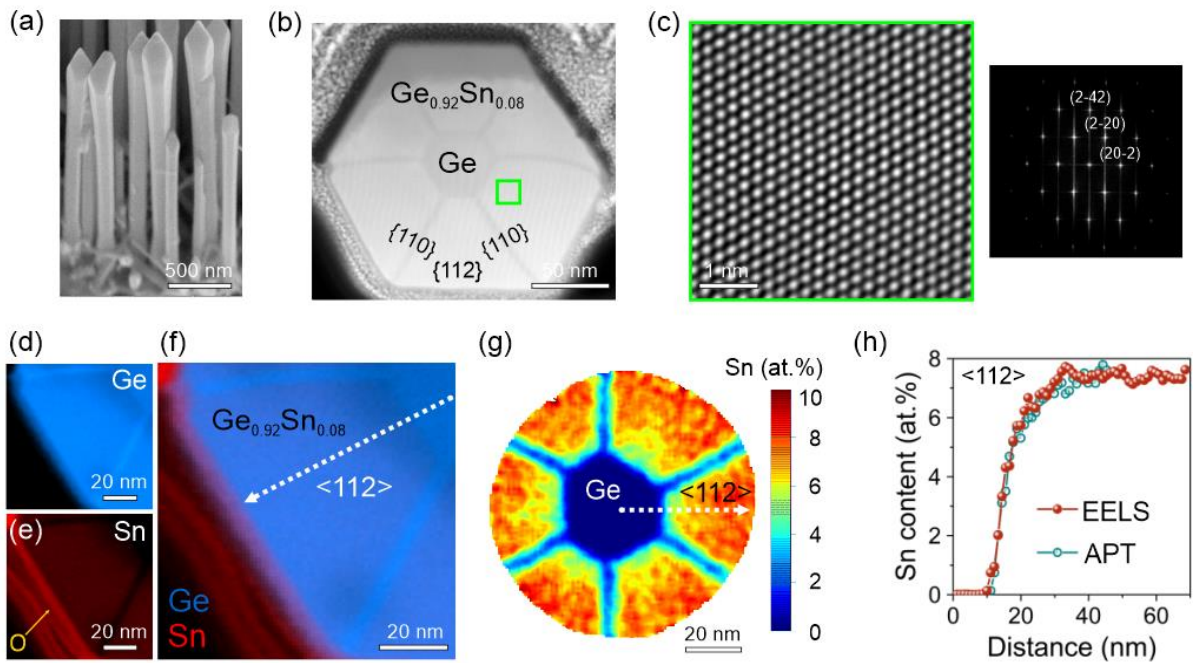

**Figure 1.** (a) SEM image of Ge/Ge$_{0.92}$Sn$_{0.08}$ core-shell NW array (b) The low magnification High Angle Annular Dark Field-Scanning Transmission Electron Microscope (HAADF-STEM) image shows the cross-section of a Ge/Ge$_{0.92}$Sn$_{0.08}$ core-shell NW with a diameter of 160 nm. (c) Atomic resolution HAADF STEM image of the defect-free Ge$_{0.92}$Sn$_{0.08}$ shell indicated by the green square in (b) and corresponding indexed FFT pattern. (d)-(e) The individual EELS maps for Ge and Sn in the NW cross section. The yellow arrow in (e) indicates the signal from O atoms in the protective oxide layer. (f) Compositional maps combined Ge and Sn atoms obtained by STEM-EELS. (g) APT measurements showing the Ge core and the inner portion of the Ge$_{0.92}$Sn$_{0.08}$ shell. The line profile (dashed arrow) is shown in (h). (h) The plot of the Sn content as a function of the distance along the radial direction for APT and EELS measurement.

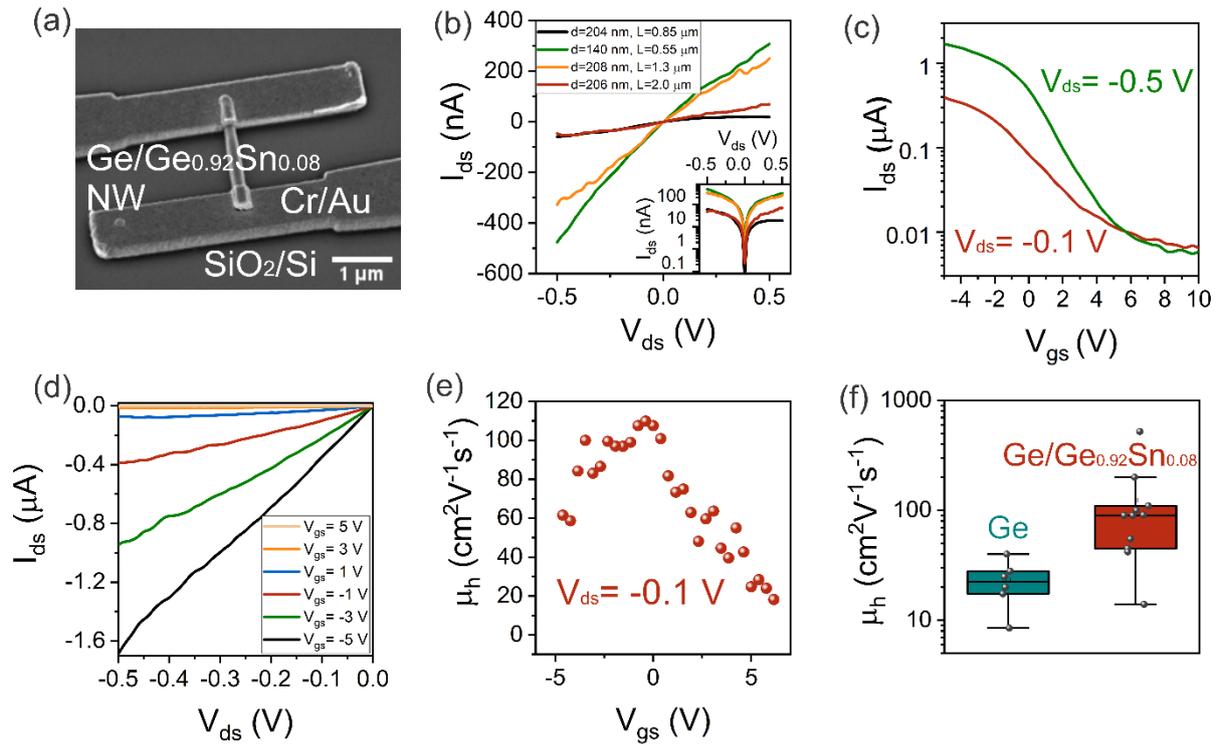

**Figure 2.** (a) SEM image of the back-gated Ge/Ge$_{0.92}$Sn$_{0.08}$ core/shell NW FET with Cr/Au metal contacts. (b) Dark current of several NWs with different diameters and channel length (c)-(d) Transfer and output characteristics of a representative Ge/Ge$_{0.92}$Sn$_{0.08}$ core/shell nanowire FET with a diameter of 250 nm. (e) Mobility assessment of the same core/shell NW FET under the source-drain bias of -0.1 V. (f) Comparing hole mobility of Ge/Ge$_{0.92}$Sn$_{0.08}$ core/shell NW and pure Ge NW FETs with different diameters.

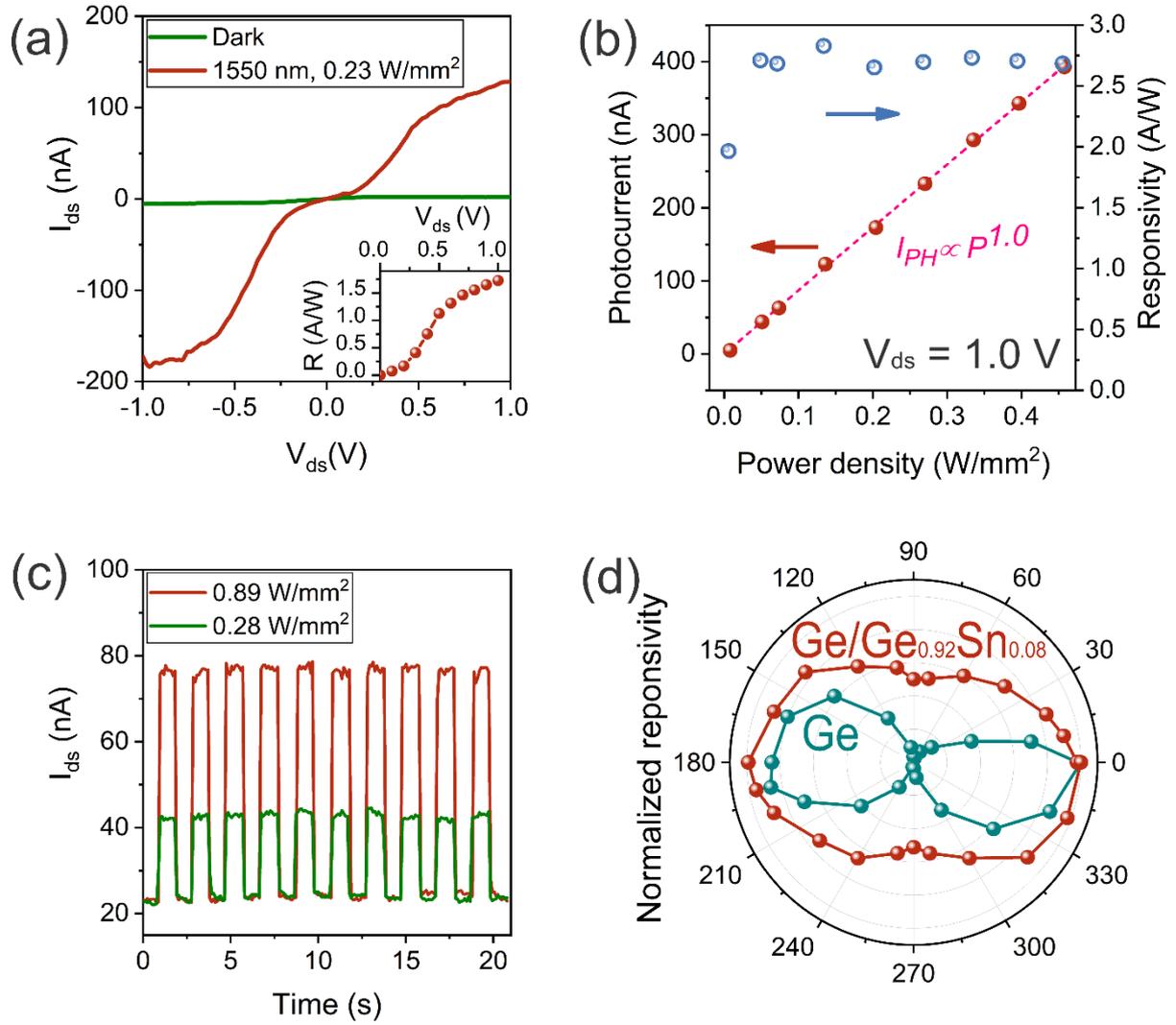

**Figure 3.** Optical characteristics of single Ge/Ge$_{0.92}$Sn$_{0.08}$ core/shell NW detector. (a) Dark current and photocurrent under 1550 nm laser with an illumination intensity of 0.23 W/mm$^2$. The inset shows the relation of responsivity with drain-source bias. (b) Dependence of the photocurrent and responsivity on light intensities. (c) Time-dependent photocurrent response of Ge/Ge$_{0.92}$Sn$_{0.08}$ core-shell NW device under 1550 nm laser with different light intensities at $V_{ds} = 0.1\ V$. (d) Polarization angle-dependent photocurrent performance of single Ge/Ge$_{0.92}$Sn$_{0.08}$ core/shell and Ge NW detector. The diameter of the core/shell NW in (a)-(c) is 190 nm, while the diameter of the core/shell NW in (d) is 240 nm.

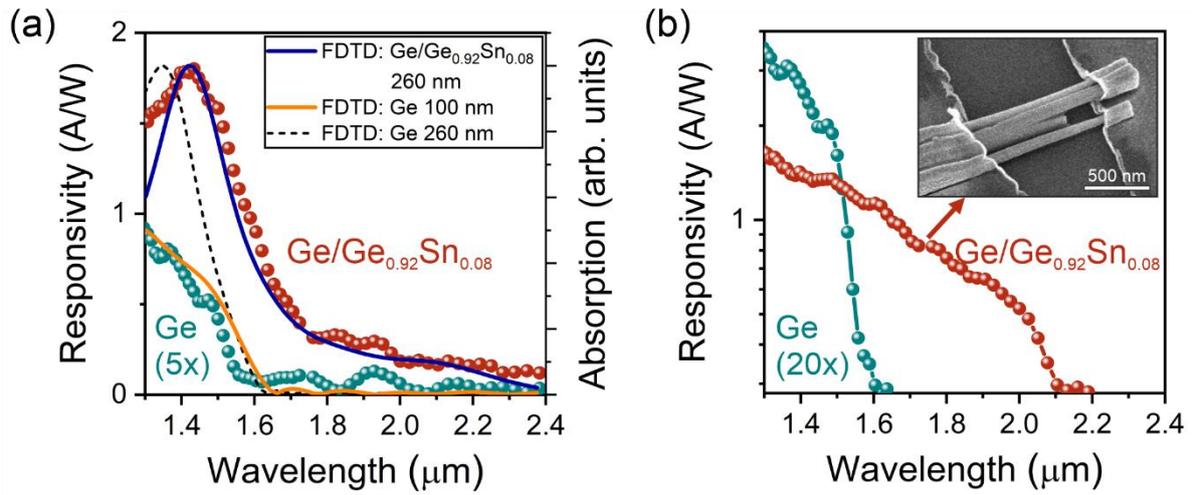

**Figure 4.** Photocurrent spectra at room temperature. (a) Responsivity measurement (spheres) and FDTD simulated absorption (solid line and dashed line) of single Ge and Ge/Ge$_{0.92}$Sn$_{0.08}$ core/shell NWs. (b) The spectral responsivity of multiple Ge/Ge$_{0.92}$Sn$_{0.08}$ core/shell and Ge NWs at 0.5 V.

**Table 1.** Electrical and optical parameters of Ge/Ge$_{0.92}$Sn$_{0.08}$ NW devices in the current work compared to Ge$_{1-x}$Sn$_x$ based nanoscale and thin-film devices reported to date.

| | This work | Ref. 51 | Ref. 21 | Ref. 52 | Ref. 49 |
|---|---|---|---|---|---|
| Sn content | 8% | 10% | 9% | 3% | 10.5 % |
| Device Structure | NW | NW | NW | Tri-gate | Thin film |
| Substrate | Si (111) | Graphene/ Ge | Si (001) | $SiO_2$ | Ge-VS |
| Growth method | CVD @ 310°C | MBE @210°C | LICVD <440°C | Polycrystalline Sub-300°C | CVD @ 310°C |
| Peak $\mu_h$ (cm$^2$V$^{-1}$s$^{-1}$) | 100 | - | 15 | 31 | - |
| Dark current (A) | 10-800 nA @ 1 V | 40 nA @ 10 mV | 10-150 nA @ 1 V | - | 2.5 mA @ 1 V |
| $I_{on}/I_{off}$ ratio | $10^2$ | - | $7.8 \times 10^2$ | $10^5$ | - |
| Responsivity @ 1.55 µ$m$ (A/W) | 2.7 @ 1 $V$ | $1.2 \times 10^{-3}$ @ 1 mV | - | - | $4 \times 10^{-3}$ @ 0.16 kV/cm |
| Cutoff (µ$m$) | 2.1 | 2.2 | - | - | 2.6 |

# Supporting Information:
# Extended Infrared Photodetection in All-Group IV Core/Shell Nanowires

Lu Luo,[†] Simone Assali,[†] Mahmoud R. M. Atalla,[†] Sebastian Koelling,[†] Anis Attiaoui,[†] Gérard Daligou,[†] Sara Martí,[‡] J. Arbiol,[‡,§]  and   Oussama Moutanabbir*,[†]

[†] Department of Engineering Physics, École Polytechnique de Montréal, C. P. 6079, Succ. Centre-Ville, Montréal, Québec H3C 3A7, Canada
[‡] Catalan Institute of Nanoscience and Nanotechnology (ICN2), CSIC and BIST, Campus UAB, Bellaterra, 08193 Barcelona, Catalonia, Spain
[§] ICREA, Pg. Lluís Companys 23, 08010 Barcelona, Catalonia, Spain

**Contents**

## S1. AC HAADF STEM images of Ge/Ge$_{0.92}$Sn$_{0.08}$ core/shell NWs

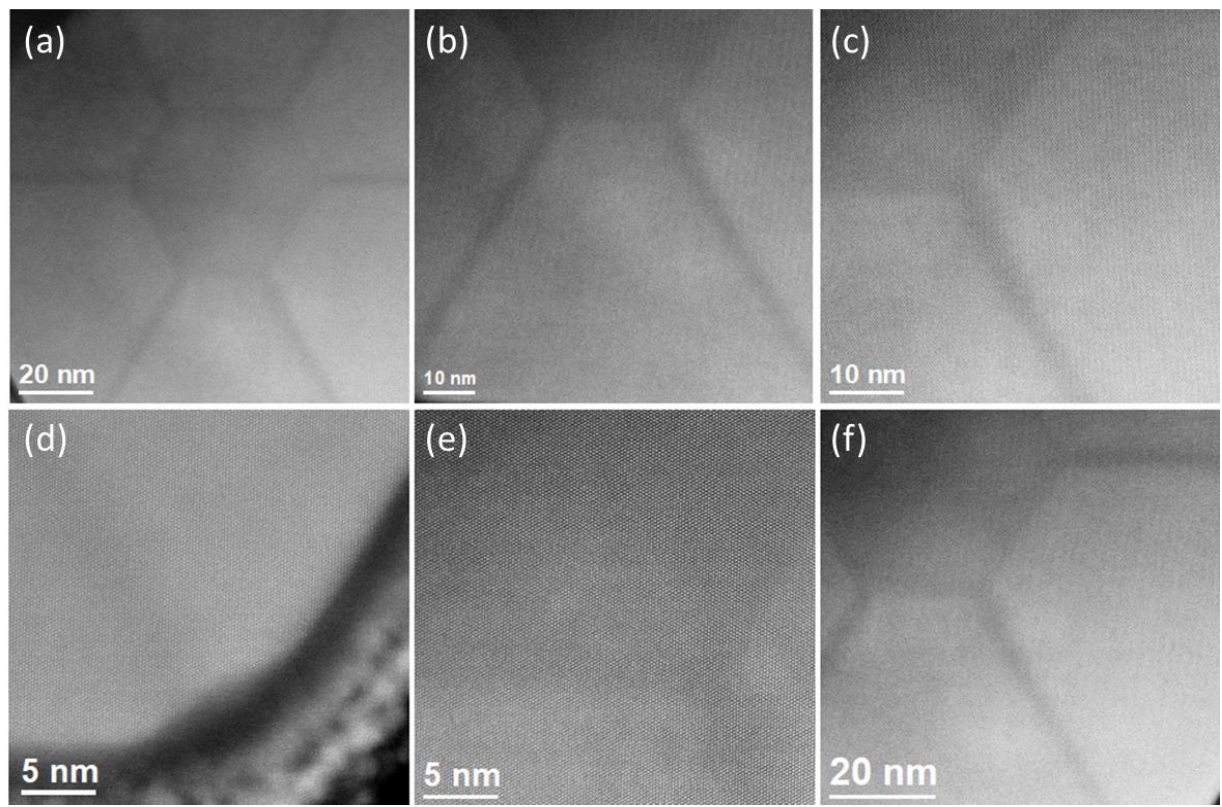

**Figure S1. (a)-(f)** Atomic resolution AC HAADF STEM images of the same Ge/Ge$_{0.92}$Sn$_{0.08}$ core/shell NW as in Fig. 1b.

## S2. Estimate the gate capacitance of NW field-effect transistor

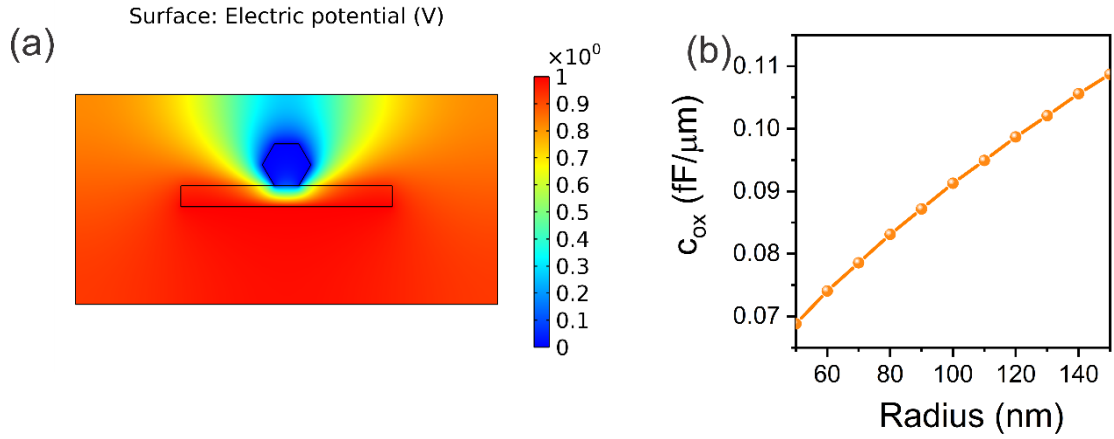

**Figure S2.** (a) 2D electric potential contour of the NW back-gated FET at zero gate bias, simulated by COMSOL software. (b) Simulated $C_{ox}$ as a function of radius per unit of the channel length.

In this work, COMSOL Multiphysics 5.6 is utilized to simulate the gate oxide capacitance ($C_{ox}$). The 2D electric potential contour of the corresponding NW back-gated FET at zero gate bias is shown in Figure S2 a. The simulated gate capacitance per unit length ($c_{ox}$) as a function of the NW radius is shown in Figure S2 b. Then the gate capacitance of a predefined NW diameter can be calculated by the equation: [32]

$$C_{ox} = c_{ox} L \qquad \text{Eq. (1)}$$

Where L is the channel length of the NW.

## S3. Electrical characteristics of single Ge field-effect transistor

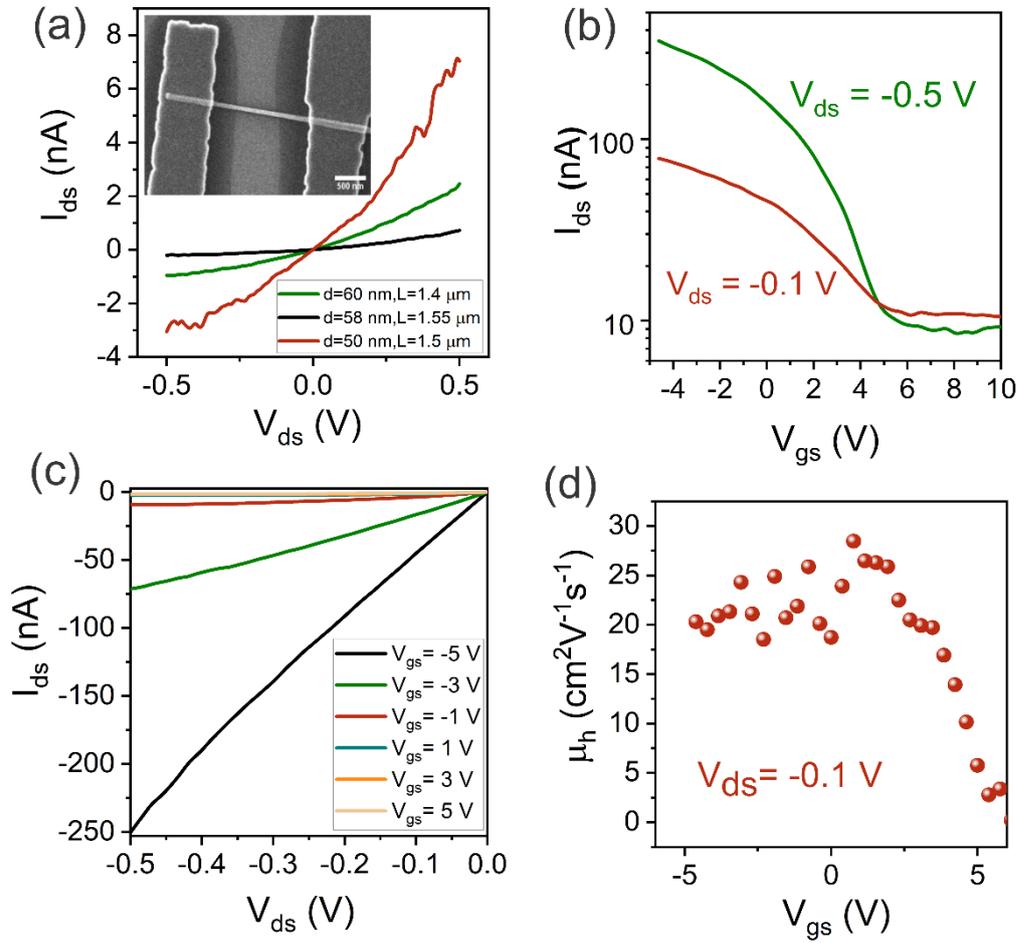

**Figure S3.** Electrical characteristics of single Ge NW FET. (a) Dark current of several single Ge NWs. Inset shows the SEM image of the Ge NW FET. (b)-(c) Transfer and output characteristics of a representative Ge nanowire FET. (d) Mobility assessment of the Ge NW FET under source-drain bias of -0.1 V.

## S4. Optical characteristics of single Ge/Ge$_{0.92}$Sn$_{0.08}$ core/shell NW detector

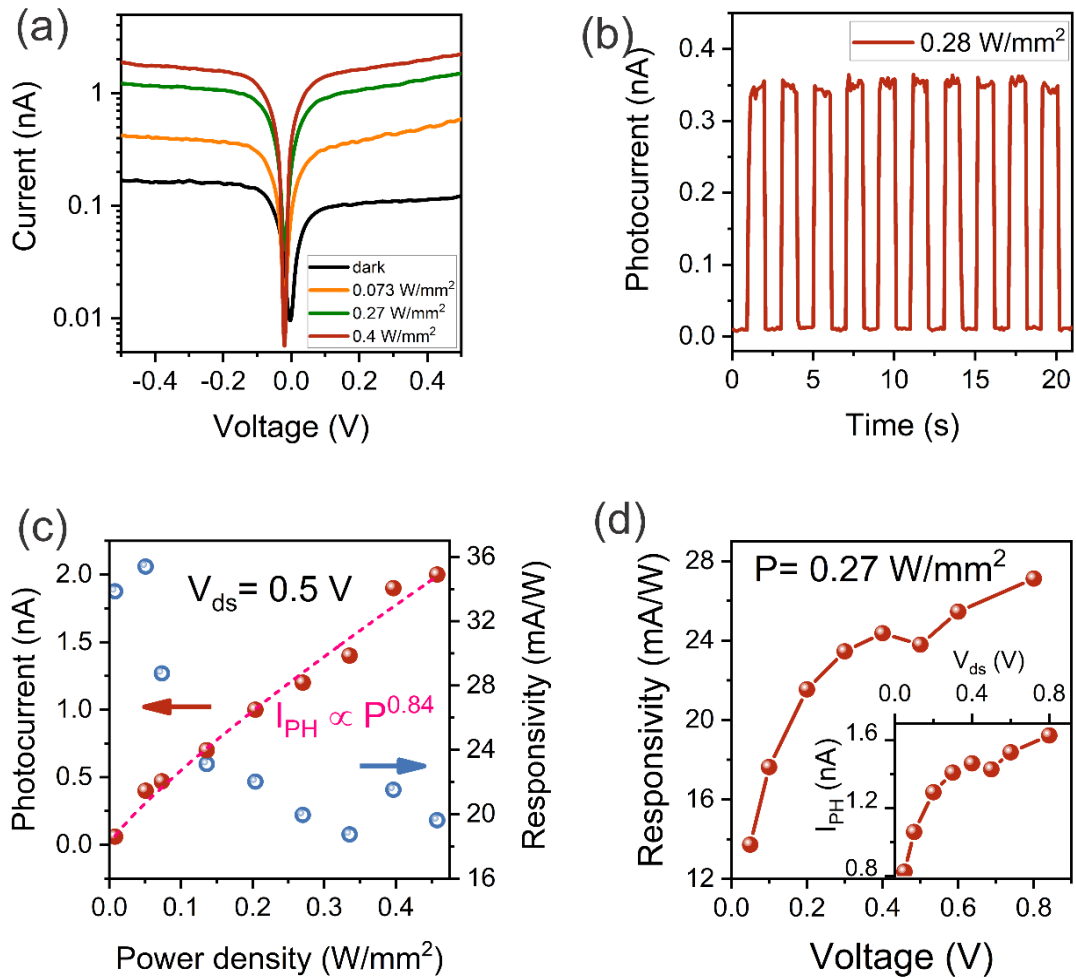

**Figure S4:** Optical characteristics of single Ge NW detector. (a) $I_{ds} - V_{ds}$ characteristics of the Ge NW photodetector under dark and 1550 nm illumination with different light intensities. (b) Photocurrent response of Ge NW device under 1550 $nm$ laser at 0.5 V. (c) Dependence of the photocurrent and responsivity on light intensity. (d) Responsivity and photocurrent as a function of the drain-source bias under a fixed power intensity of 0.27 $W/mm^2$.

## S5. FDTD simulation

Absorption efficiency ($Q_{abs}$), defined as the ratio of absorption cross-section and the physical cross-section of the NW wire, is simulated by using Lumerical FDTD © software. The 2D model is depicted in Figure S5. The diameter of Ge core and the thickness of the $Ge_{0.92}Sn_{0.08}$ shell of the NW are 60 nm and 100 nm, respectively. The Total-Field Scattered-Field (TFSF) source plane is incident perpendicular to the NW, and the perfect matching layer (PML) is selected as the simulation boundary. The optical properties of $Ge_{0.91}Sn_{0.09}$ were used in the simulation, but the change of the optical parameters of the $Ge_{0.92}Sn_{0.08}$ and $Ge_{0.91}Sn_{0.09}$ is negligible and won't affect the simulations. The optical properties of Ge and $Ge_{0.91}Sn_{0.09}$ were extracted from spectroscopic ellipsometry measurement. Monitor for calculating absorption cross-section of the NW is indicated by the yellow square. In order to correspond to the experimental absorption spectrum under unpolarized light, the simulated unpolarized absorption efficiency is calculated by

$$Q_{abs} = \frac{Q_{abs}^{TE} + Q_{abs}^{TM}}{2} \qquad \text{Eq. (2)}$$

where $Q_{abs}^{TE}$ and $Q_{abs}^{TM}$ are TE and TM polarized absorption efficiency of the NWs.[42]

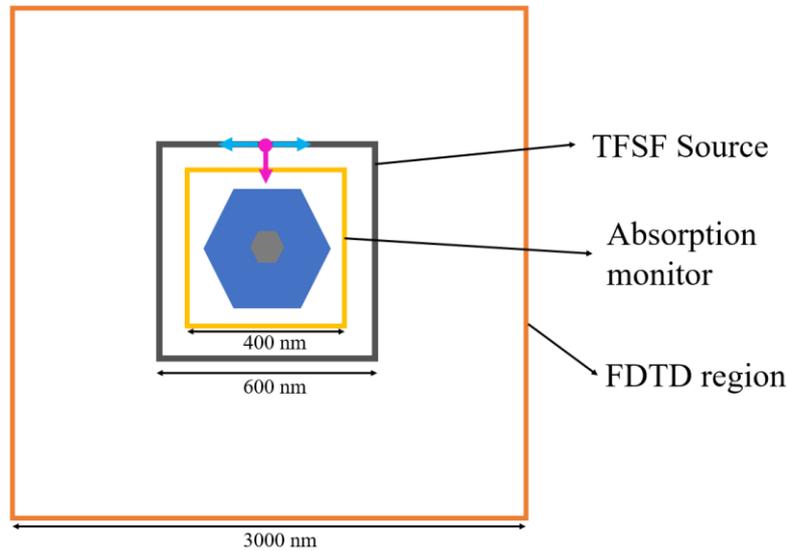

**Figure S5:** $Ge/Ge_{0.92}Sn_{0.08}$ core/shell NW 2D model.